\documentclass[aps,prb,twocolumn,superscriptaddress]{revtex4}
\usepackage{graphicx}
\usepackage{amsmath}

\begin{document}

\title{Coulomb Blockade of a~Three-terminal Quantum Dot}

\author{Robert Andrzej \.Zak}
\affiliation{Institute of Theoretical Physics, Faculty of Physics,
University of Warsaw, ul. Ho\.za 69, PL-00-681 Warsaw, Poland}
\affiliation{Nano-Science Center, Niels Bohr Institute, University
of Copenhagen, Universitetsparken 5, DK-2100 Copenhagen \O, Denmark}

\author{Karsten Flensberg}
\affiliation{Nano-Science Center, Niels Bohr Institute, University
of Copenhagen, Universitetsparken 5, DK-2100 Copenhagen \O, Denmark}

\begin{abstract}
We study an interacting single-level quantum dot weakly coupled to
three electrodes. When two electrodes are biased by voltages with
opposite polarities, while keeping the third lead (the stem)
grounded, the current through the stem is a measure of electron-hole
asymmetry of the dot. In~this setup we calculate the stem current
for both metallic and ferromagnetic (collinearly polarized) leads
and discuss how the three-terminal device gives additional
information compared to the usual two-terminal setup. We calculate
both the sequential and cotunneling contribution for the currents.
For the latter part we include a regularization procedure for the
cotunneling current, which enables us to also describe the behavior
at the charge degeneracy points.
\end{abstract}


\maketitle

\section{\label{sec:int}INTRODUCTION}

A systems consisting of a~quantum dot weakly coupled to external
leads have been extensively studied, both for
unpolarized\cite{PhysRevB.54.16820,Lehm06,koch:056803} and polarized
leads.\cite{PhysRevB.62.1186,PhysRevB.64.085318,weymann:115334,
PhysRevLett.90.166602,braun:195345,Ped05} These all dealt with
two-terminal transport properties, with a third electrode acting as
a capacitively connected gate.

System with more than two terminals has a long history in mesoscopic
physics.\cite{buttiker} The functionality added by the third lead is
of importance for some applications, in particular the so-called
Y-branch structures.\cite{Palm92,
PhysRevLett.82.2564,PhysRevB.62.16727,Worchech05} These studies
consider open mesoscopic systems, where correlation effects can be
neglected. Some recent works have studied the opposite case where
correlations are important, \emph{e.g.} the study of current-current
correlations in a~three-terminal
device,\cite{cottet:115315,cottet:206801} Kondo peaks in a
three/four-terminal setup,\cite{PhysRevB.64.153306,sanchezlopez} and
crossed carbon nanotubes.\cite{PhysRevLett.89.226404,
Egger03,gao:216804} Here, we consider a multi-terminal quantum dot
and, in particular, the non-linear response, which brings in new
information about the structure of the quantum dot as well as about
the magnetization of side branches.

\begin{figure}
\includegraphics[width=.4\textwidth]{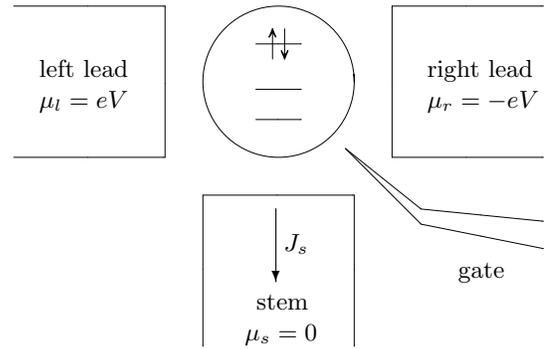}
\caption{\label{fig:3TermDev}The three terminal device. The
occupation number is controlled via the gate by applying the voltage
$V_{g}$. The stem chemical potential, $\mu_{s}=0$, serves as the
reference level. The left and right leads are biased in push-pull
manner with $\mu_{l}=eV$ and $\mu_{r}=-eV$.}
\end{figure}

The system we study in detail consists of a~single level quantum dot
coupled to three leads, Fig.~\ref{fig:3TermDev}. We allow collinear
spin polarization of leads, so that some of them can be polarized
while others are kept unpolarized. For the sake of simplicity, the
polarization, if any, is assumed to be complete. To lift the spin
degeneracy, we apply a~magnetic field to the dot collinear with
polarization of the leads. We study the current through the central
junction (referred to as a~stem) and its dependence on the applied
voltage, magnetic field, position of dot's energy states, and
orientation of leads' polarization. Except for the limiting case of
non-interacting electrons, the exact solution is not known, and
approximative methods are widely applied. In the typical
experimental setup a~coupling between the dot and the lead is of
order of $\mu$eV while temperature of order of meV or higher,
therefore it is often justified to perform perturbation expansion in
a~small parameter $\Gamma/k_{B}T$. We cut off the perturbation
expansion on second order.

In first order, we derive the sequential tunneling current  through a
central electrode, Eq.~(\ref{eq:JsSeq}). When the system is biased
symmetrically, $V_{l}=V$ and $V_{r}=-V$, the stem current is an~even
function of applied voltage and an~odd function of the gate voltage,
Fig.~\ref{fig:Dens}. The stem current can be either negative or
positive depending on the nature of the electron transport.

To understand this, we first observe that the current from the left
to the right lead occurs at low temperatures via only two charge
states and can be predominately either electron- or hole-like. If
the energy difference between the $\nu+1$ and $\nu$ electron states
is positive, the transport is electron-like and hole-like if it is
negative, with $\nu$ standing for the occupation number. For the
electron-like case the $\nu+1$ state can decay via an~electron
leaving through the stem, whereas for the hole-like case, the $\nu$
state can decay by the electron entering from the stem.
\textit{Hence the stem current is zero for the electron-hole
symmetric case.}

The three terminal setup therefore measures the electron-hole
asymmetry, and in this respect it is similar to
thermopower.\cite{Matv02} Interestingly, a~change between the two
situations above, can be induced by the magnetic field applied to
the dot which modifies the physics significantly, see
Fig.~\ref{fig:JsvsB}. Moreover, in a~three-terminal setup with
magnetized side branches and the non-magnetic stem the current value
enables one to distinguish among four magnetization alignments,
Fig.~\ref{fig:DensB}.

\begin{figure}
\includegraphics[width=.45\textwidth]{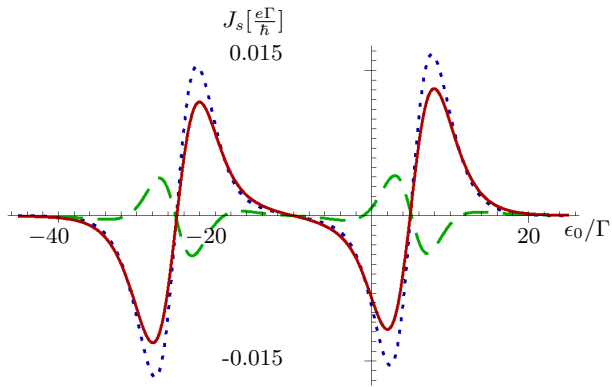}
\caption{\label{fig:Jtot}(Color online)
    The stem current (solid, red) being the sum of the sequential part (dotted, blue)
    and the cotunneling (dashed, green). The sequential contribution
    gives sufficient qualitative description. For this plot leads are
    unpolarized with $\Gamma_{l}=\Gamma_{r}=\Gamma_{s}=\Gamma/3$, $U=20\Gamma$,
    $B=5\Gamma$, $k_{B}T=2\Gamma$ and $V/\Gamma=1$.}
\end{figure}

In second order, the total current is increased by virtual
cotunneling processes for these values of the gate voltage for which
the sequential tunneling is exponentially suppressed. Furthermore,
the cotunneling correction lowers the sequential current maxima and
gives rise to additional broadening of order of the sum of all
couplings $\Gamma_{s}+\Gamma_{l}+\Gamma_{r}$, and hence the total
broadening is the sum of the thermal part and the latter. However,
we will show that the cotunneling current does not change the
sensitivity towards electron-hole asymmetry, and therefore the
qualitative picture described above still holds to higher order in
tunneling, see Fig.~\ref{fig:Jtot}.

In the Appendix the exact solution for non-interacting electrons is
compared with the perturbative treatment and even though the
renormalization of level positions also contributing to the second
order current was neglected, we achieve good agreement between the
two.

\section{\label{sec:mod}MODEL HAMILTONIAN}

The model Hamiltonian of the quantum dot connected to any number of
unpolarized leads is
\begin{equation}\label{eq:GeneralQDHam}
    H=H_{L}+H_{T}+H_{D},
\end{equation}
where
\begin{equation}\label{eq:LeadHam}
    H_{L}=\sum_{j}H_{L}^{j}=\sum_{j}\sum_{k\sigma}
    \xi _{j,k\sigma}c_{j,k\sigma}^{\dagger}c_{j,k\sigma}
\end{equation}
describes the uncoupled leads indexed by $j$ while
$c_{j,k\sigma}^{\dagger}$ and $c_{j,k\sigma }$ form an~orthogonal
set of creation and annihilation operators in lead $j$.

The dot is described by the Hamiltonian
\begin{equation}\label{eq:QDHam}
    H_{D}=\sum_{\sigma }(\epsilon_{0}-\sigma B)d_{\sigma}^{\dagger}d_{\sigma}
    +Un_{\uparrow}n_{\downarrow}
\end{equation}
with a~single electron of the orbital energy $\epsilon_{0}$, the
Coulomb repulsion energy $U$, magnetic field $B$, $\sigma=1$ for
a~spin-up electron and $\sigma=-1$ for a~spin-down electron,
$d_{\sigma}^{\dagger}$ and $d_{\sigma}$ forming a~set of orthogonal
creation and annihilation operators for the dot and
$n_{\sigma}=d_{\sigma}^{\dagger}d_{\sigma}$ being an~occupation
number operator. The tunneling processes between the leads and the
central region are taken into account via the tunneling Hamiltonian
\begin{equation}\label{eq:TunHam}
    H_{T}=\sum_{j}\sum_{k\sigma}(t_{j,k\sigma}
    c_{j,k\sigma}^{\dagger}d_{\sigma}+\textrm{h.c.}),
\end{equation}
where $t_{j,k\sigma}$ is a~spin-dependent tunneling amplitude.

To include the complete magnetization of one of the leads, say lead
$j$, one should limit the corresponding Hilbert space to one spin
direction replacing $\sigma$ by $\uparrow$ or $\downarrow$.

\section{\label{sec:seq}SEQUENTIAL TUNNELING REGIME}

The simplest situation we study is the sequential tunneling regime,
also called the weak tunneling regime. It is assumed that the time
between tunneling events is the largest time scale in the problem,
so that there is no coherence between successive tunneling
processes. If there is no bias applied, the distribution function of
different states is given  by the equilibrium Gibbs function. With
an applied voltage difference between the electrodes the induced
non-equilibrium distribution function needs to be determined.

To this end, we calculate the transition rates between different
dot's states. Since we consider the weak tunneling regime, the
Fermi's Golden Rule is sufficient to tackle this problem. We define
the transition rates $\Gamma^j_{\mu\nu}$ as the rate for a~process
that changes the state of the dot from $\nu$ to $\mu$ due to
tunneling through junction $j$. Assuming a~continuous density of
states inside the electrodes it follows that the tunneling rates are
proportionate to the Fermi function,
$\Gamma_{\nu+1,\nu}^{j}=\Gamma^{0}_{j}
n(\epsilon_{\nu+1}-\epsilon_{\nu}-\mu_{j})$, if an~electron jumps
onto the dot, and $\Gamma_{\nu-1,\nu}^{j}=\Gamma^{0}_{j}
\big(1-n(\epsilon_{\nu-1}-\epsilon_{\nu}-\mu_{j})\big)$ for the
opposite process. Here $\mu_{j}$ is a~chemical potential of
reservoir~$j$, $\Gamma^{0}_{j}\equiv2\pi|t_{j}|^{2}\rho_{j}$ with
$\rho_{i}$ being the density of states in lead $j$. To include the
collinear polarizations $P_{i}$ of the leads spin--dependent
tunneling rates
$\Gamma_{i\sigma}^{0}=\frac{1}{2}\Gamma_{i}^{0}(1+\sigma P_{i})$ are
introduced.

Having found the transitions rates, we can write down master
equations describing the dynamical behavior of the distribution
function $P_{\nu}$ and since we are only interested in the steady
state solution, we have
\begin{widetext}
\begin{equation}\label{eq:Master}
\left(
    \begin{array}{cccc}
        -\Gamma_{\uparrow0}-\Gamma_{\downarrow0} & \Gamma_{0\uparrow} & \Gamma_{0\downarrow} & 0 \\
        \Gamma_{\uparrow0} & -\Gamma_{2\uparrow}-\Gamma_{0\uparrow} & 0 & \Gamma_{\uparrow2} \\
        \Gamma_{\downarrow 0} & 0 & -\Gamma_{2\downarrow}-\Gamma_{0\downarrow} & \Gamma_{\downarrow2} \\
        0 & \Gamma_{2\uparrow} & \Gamma_{2\downarrow} & -\Gamma_{\uparrow2}-\Gamma_{\downarrow2} \\
    \end{array}
    \right)
    \left(
    \begin{array}{c}
        P_{0} \\
        P_{\uparrow} \\
        P_{\downarrow} \\
        P_{2} \\
    \end{array}
    \right)=0,
\end{equation}
\end{widetext}
where $\Gamma_{\mu\nu}=\sum_{j}\Gamma^{j}_{\mu\nu}$ is a~total
contribution from all the leads to the transition rate from
state~$\nu$ to state~$\mu$, while $P_{0}$, $P_{\uparrow}$,
$P_{\downarrow}$, $P_{2}$ are the empty, spin-up, spin-down and
double occupation state distribution functions respectively. The
terms with a~minus sign give the rate at which a~given state on the
left-hand side decays while the plus--sign terms describe the
opposite processes. In addition to these equations, the probability
conservation law $P_{0}+P_{\uparrow}+P_{\downarrow}+P_{2}=1$ has to
be used.

The knowledge of the distribution functions allows one to find the
net current flowing through junction $i$
\begin{equation}\label{eq:JiSeq}
    J_{i}=-\frac{e}{\hbar}\sum_{\nu}(\Gamma_{\nu+1,\nu}^{i}
    -\Gamma_{\nu-1,\nu}^{i})P_{\nu}.
\end{equation}
This should be understood as a~difference between the number of
electrons incoming to the dot and deoccupating it, times the
electron charge, $-e<0$.

\emph{We choose the Fermi level of the stem reservoir to be our
reference level, $\mu_{s}=0$, throughout the text and in all
diagrams}.

The solutions to Eqs.~(\ref{eq:Master}) are
\begin{subequations}\label{eq:MasterSol}
\begin{align}
    \hspace{-5pt}P_{0}=\frac{1}{D}\big(\Gamma_{0\uparrow}\Gamma_{\uparrow 2}
    (\Gamma_{0\downarrow}+\Gamma_{2\downarrow})
    +\Gamma_{0\downarrow}\Gamma_{\downarrow 2}
    (\Gamma_{0\uparrow}+\Gamma_{2\uparrow})\big),\\
    \hspace{-5pt}P_{\uparrow}=\frac{1}{D}\big(\Gamma_{\uparrow0}\Gamma_{0\downarrow}
    (\Gamma_{\downarrow 2}+\Gamma_{\uparrow 2})
    +\Gamma_{\uparrow 2}\Gamma_{2\downarrow }
    (\Gamma_{\downarrow0}+\Gamma_{\uparrow0})\big),\\
    \hspace{-5pt}P_{\downarrow}=\frac{1}{D}\big(\Gamma_{\downarrow0}
    \Gamma_{0\uparrow}(\Gamma_{\downarrow 2}+\Gamma_{\uparrow 2})
    +\Gamma_{\downarrow 2}\Gamma_{2\uparrow }
    (\Gamma_{\downarrow0}+\Gamma_{\uparrow0})\big),\\
    \hspace{-5pt}P_{2}=\frac{1}{D}\big(\Gamma_{2\uparrow}\Gamma_{\uparrow 0}
    (\Gamma_{0\downarrow}+\Gamma_{2\downarrow})
    +\Gamma_{2\downarrow}\Gamma_{\downarrow0}
    (\Gamma_{0\uparrow}+\Gamma_{2\uparrow})\big),
\end{align}
\end{subequations}
with
\begin{align}
    \notag D&=(\Gamma_{\downarrow 0}\Gamma_{0\uparrow}
    +\Gamma_{2\uparrow}\Gamma_{\uparrow 0})
    (\Gamma_{\downarrow 2}+\Gamma_{2\downarrow})\\
    \notag &+(\Gamma_{0\downarrow}\Gamma_{\downarrow 2}
    +\Gamma_{\uparrow 2}\Gamma_{2\downarrow})
    (\Gamma_{\uparrow0}+\Gamma_{0\uparrow}) \notag \\
    \notag &+(\Gamma_{0\uparrow}\Gamma_{\uparrow 2}
    +\Gamma_{\downarrow 2}\Gamma_{2\uparrow})
    (\Gamma_{\downarrow 0}+\Gamma_{0\downarrow})\\
    &+(\Gamma_{\uparrow 0}\Gamma_{0\downarrow}
    +\Gamma_{2\downarrow}\Gamma_{\downarrow 0})
    (\Gamma_{\uparrow 2}+\Gamma_{2\uparrow}).
\end{align}

Using Eq.~(\ref{eq:JiSeq}) the stem current becomes
\begin{align}\label{eq:JsSeq}
    J_{s}=-\frac{e}{\hbar}
    \big(&(\Gamma_{\uparrow0}^{s}+\Gamma_{\downarrow0}^{s})P_{0}
    +(\Gamma_{2\uparrow}^{s}-\Gamma_{0\uparrow}^{s})P_{\uparrow}  \notag \\
    +&(\Gamma_{2\downarrow}^{s}-\Gamma_{0\downarrow}^{s})P_{\downarrow}
    -(\Gamma_{\downarrow 2}^{s}+\Gamma_{\uparrow 2}^{s})P_{2}\big).
\end{align}
The above expression is a~complicated combination of the Fermi
functions and analytic treatment is not further possible. We discuss
numerical results in Sec.~\ref{sec:sum}.

\section{\label{sec:cot}COTUNNELING REGIME}

As discussed in the introduction, one could expect the cotunneling
current to modify the lowest order current significantly. However,
this turns out not to be the case, since the cotunneling current
also has nodes in the particle-hole symmetric points and only give
small corrections to the overall shape of the current versus gate or
voltage curves.

In the cotunneling regime two electron processes come into play. An
electron is transferred between lead $j$ and lead $i$ (via an
intermediate classically forbidden state) in two successive
tunneling events, across the quantum dot. For the calculation of the
two-electron rates,  we need to consider the different starting
configurations (empty, single- and double-occupied). The
probabilities that a~given state is occupied are given by
Eqs.~(\ref{eq:MasterSol}). Furthermore, there are two types of
processes that should be considered, \emph{i.e.} non-spin-flip
processes that do not change dot's magnetization, and spin-flip
processes leading to reversal of spin direction, see
Fig.~\ref{fig:SpinFlip}. The latter contribute directly to the
current as well as modify the probabilities $P_{\sigma}$ via
spin-flips caused by the interaction of the dot with the lead. We
find the rates for these processes employing the generalized Fermi's
Golden Rule
\begin{equation}
    \Gamma_{fi}=\frac{2\pi}{\hbar}\big|\langle f|T|i\rangle\big|^{2}
    \delta(E_{f}-E_{i})
\end{equation}
for the transition from the initial state $|i\rangle$ of energy
$E_{i}$ to the final state $|f\rangle$ of energy $E_{f}$. The
transition operator $T$ is defined as
\begin{equation}
    T=H_{T}+H_{T}\frac{1}{E_{i}-H_{0}}T
\end{equation}
with $H_{0}=H_{L}+H_{D}$. The first non-vanishing transition rate
for the process which transfers the electron between the electrodes
appears in the perturbation expansion in second order, thus the
cotunneling process is quadratic in couplings, $\Gamma$.

The cotunneling events are two-particle processes and therefore
occur only between pairs of the leads. Hence, the cotunneling
current through junction $i$ can be expressed as the sum of the
currents between any pair of leads $\tilde{J}^{\nu}_{ij}$,
\emph{i.e.} the currents between lead $i$ and $j$ in regime~$\nu$,
weighted by the appropriate probability $P_{\nu}$
\begin{equation}\label{eq:CurCotTot}
\index{Cotunneling current}
    \tilde{J}_{i}=\sum_{j}\Big(\sum_{\nu}P_{\nu}\tilde{J}^{\nu}_{ij}
    +\sum_{\sigma}P_{\sigma}\tilde{J}^{\sigma,sf}_{ij}\Big).
\end{equation}
The sum runs over dot's states $\nu=0,\uparrow,\downarrow,2$, and
for emphasis we divided the current into the non-spin-flip part
$\tilde{J}^{\nu}_{ij}$ and the spin-flip part
$\tilde{J}^{\sigma,sf}_{ij}$ (present for the single occupation
only).

\subsection{\label{sec:cotnsf}Non-spin-flip cotunneling current}

We exemplify the derivation of the current between a~pair of leads
by the case of the empty dot. The initial state
$|i\rangle=|\nu_{1},\dots,\nu_{N},0\rangle$ consists of a~tensor
product of lead's states $|\nu_{i}\rangle$ and the dot's state
$|0\rangle$. The electron can be transferred from lead $i$ into lead
$j$ via either a~spin-up or spin-down state of the dot, depending on
a~spin of the electron entering the intermediate region. Because the
two corresponding final states of the leads are different
$|f\rangle=|\nu_{1},\dots,\nu_{i}-\sigma,\dots,\nu_{j}+\sigma,\dots,\nu_{N},0\rangle
=c_{i,k\sigma}c^{\dagger}_{j,k'\sigma}|i\rangle$, there is no
interference between electron's paths and we find the rates
$\Gamma^{0}_{j\sigma i\sigma}$ for these two processes separately
and add them up to get the total tunneling rate in this regime.
Thus, the current between the pair of leads $i,j$ is given by
\begin{equation}\label{eq:PairedCur0Def}
    \tilde{J}^{0}_{ij}=-e\sum_{\sigma}(\Gamma^{0}_{j\sigma i\sigma}
    -\Gamma^{0}_{i\sigma j\sigma}),
\end{equation}
that is the rate for the process bringing the electron from lead~$i$
to lead~$j$ through the empty state, minus the rate for the opposite
process, multiplied by the electron charge, $-e$. We substitute the
tunneling Hamiltonian $H_{T}$, Eq.~(\ref{eq:TunHam}), into $\langle
f|T|i\rangle$ and after a number of standard calculations, where we
use \textit{(i)} that the distribution function of electron states
in the leads are given by the Fermi-Dirac distribution functions and
\textit{(ii)} the assumption of a constant tunneling density of
states, we arrive at
\begin{equation}
    \Gamma^{0}_{j\sigma i\sigma}=\frac{1}{h}
    \Gamma^{0}_{i\sigma}\Gamma^{0}_{j\sigma}
    \int_{-\infty}^{\infty}d\xi
    \frac{1}{(\xi-\epsilon_{\sigma})^{2}}
    n(\xi-\mu_{i})\big(1-n(\xi-\mu_{j})\big)
\end{equation}
Employing Eq.~(\ref{eq:PairedCur0Def}), the current between lead $i$
and $j$ through the empty dot becomes
\begin{subequations}\label{eq:PairedCur}
\begin{align}
    \notag \tilde{J}_{ij}^{0}=-\frac{e}{h}\Gamma^{0}_{i}\Gamma^{0}_{j}
    \int^{\infty}_{-\infty}&d\xi\bigg(\frac{1}
    {(\xi-\epsilon_{\uparrow})^{2}}
    +\frac{1}{(\xi-\epsilon_{\downarrow})^{2}}\bigg)\\
    &\times\big(n(\xi-\mu_{i})-n(\xi-\mu_{j})\big),
\end{align}
where we took
$\Gamma^{0}_{i}=\Gamma^{0}_{i\uparrow}=\Gamma^{0}_{i\downarrow}$. If
one neglects the inelastic processes, the currents through the
single occupied dot may be found by analogy
\begin{align}\label{eq:PairedCur1}
    \notag \tilde{J}^{\sigma}_{ij}=-\frac{e}{h}\Gamma^{0}_{i}\Gamma^{0}_{j}
    \int^{\infty}_{-\infty}&d\xi
    \bigg(\frac{1}{(\xi-\epsilon_{\sigma})^{2}}
    +\frac{1}{(\xi-\epsilon_{\bar{\sigma}}-U)^{2}}\bigg)\\
    &\times\big(n(\xi-\mu_{i})-n(\xi-\mu_{j})\big)
\end{align}
with $\bar{\sigma}=-\sigma$. The expression for the current in the
presence of spin-flipping will be derived below as it needs more
attention.

Finally, the currents $J_{ij}^{2}$ in the remaining regime of the
double occupation reads
\begin{align}
    \notag \tilde{J}_{ij}^{2}=-\frac{e}{h}\Gamma^{0}_{i}\Gamma^{0}_{j}
    \int^{\infty}_{-\infty}&d\xi
    \bigg(\frac{1}{(\xi-\epsilon_{\uparrow}-U)^{2}}
    +\frac{1}{(\xi-\epsilon_{\downarrow}-U)^{2}}\bigg)\\
    &\times\big(n(\xi-\mu_{i})-n(\xi-\mu_{j})\big).
\end{align}
\end{subequations}
These expressions are, however, divergent and we need to improve on
the second order perturbation theory (in $\Gamma$) to regularize
these divergences. Before we follow the regularization procedure in
Sec.~\ref{sec:cotreg}, we study the current flowing with reversal of
the spin accumulated on the dot.

\subsection{\label{sec:cotsf}Spin-flip cotunneling current}

As long as we consider the limit of non-interacting electrons,
$U=0$, the inelastic processes do not affect the current at all, and
the above description is completely sufficient. In the Appendix we
compare the perturbative result with the exact results obtainable
when neglecting interactions.

\begin{figure}
\includegraphics[width=.4\textwidth]{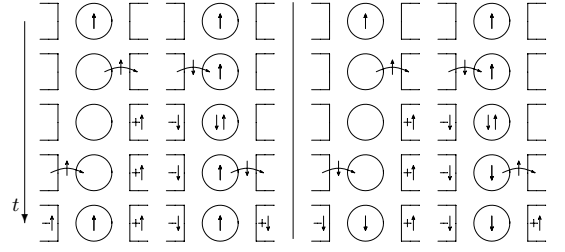}
\caption{\label{fig:SpinFlip}The comparison of
    a~cotunneling event for the dot occupied by a~spin-up electron with
    the spin-flip process (on the right), and without (on the left). In case
    of inelastic processes, the final state of the dot and the leads
    is the same for both paths resulting in the interference.}
\end{figure}

The derivation of the tunneling rate and the current between a pair
of leads in the presence of spin-flip processes is similar to one
used in the previous section. Here, we just cite the result
\begin{equation}\label{eq:PairedCurSFDef}
    \tilde{J}^{\sigma,sf}_{ij}=-e(\Gamma_{j\bar{\sigma}i\sigma}
    -\Gamma_{i\bar{\sigma}j\sigma})
\end{equation}
with the tunneling rate
\begin{align}
    \notag\Gamma_{j\bar{\sigma}i\sigma}=\frac{1}{h}
    \Gamma^{0}_{i\sigma}\Gamma^{0}_{j\bar{\sigma}}
    \int_{-\infty}^{\infty}d\xi
    &\bigg(\frac{1}{\xi-\epsilon_{\bar{\sigma}}-U}
    -\frac{1}{\xi-\epsilon_{\bar{\sigma}}}\bigg)^{2}\\
    \times&n(\xi-\mu_{i})\big(1-n(\xi-\mu_{j}
    -\Delta_{\bar{\sigma}\sigma})\big).
\end{align}
This expression differs from that for the cotunneling current
without spin-flip processes, Eq.~(\ref{eq:PairedCur1}), as the
interference term appears there, Fig.~\ref{fig:SpinFlip}.
Furthermore, the chemical potential of lead $j$ is shifted due to
the energy gap,
$\Delta_{\bar{\sigma}\sigma}\equiv\epsilon_{\bar{\sigma}}-\epsilon_{\sigma}$,
resulting from the Zeeman splitting. It is convenient to rewrite the
expression
$n(\xi-\mu_{i})\big(1-n(\xi-\mu_{j}+\Delta_{\bar{\sigma}\sigma})\big)
=n_{B}(\mu_{j}-\mu_{i}+\Delta_{\bar{\sigma}\sigma})\times
\big(n(\xi-\mu_{j}-\Delta_{\bar{\sigma}\sigma})-n(\xi-\mu_{i})\big)$
so that it has the same form as Eqs.~(\ref{eq:PairedCur}),
\emph{i.e.} it is proportional to the Fermi functions' difference,
with $n_{B}$ standing for the Boltzmann function. In the next
section it will come to light that terms of this type undergo the
regularization scheme.

\subsection{\label{sec:cotreg}Regularization procedure}

In general for the non-spin-flip cotunneling current, the
problematic integrand is a~product of the divergent term
$(\xi-\xi_{0})^{-2}$ and the Fermi functions' difference, denoted as
$f(\xi)\equiv n(\xi-\mu_{i})-n(\xi-\mu_{j})$.

To deal with this divergence we follow the regularization scheme
proposed by Turek and Matveev \cite{Matv02,vonoppen} and add to the
denominator a life-time broadening, $\eta^{2}$, describing the
tunneling broadening of the intermediate state,
\emph{i.e.}~$\eta\propto\Gamma$. The whole trick is to partition
this integral into two parts, from which the first can be
\emph{a~posteriori} identified as the energy conserving process, and
hence, the sequential tunneling contribution while the second term
describes the regularized cotunneling processes. This is done as
\begin{align}\label{eq:PartSeqCo}
    \notag\int d\xi&\frac{f(\xi)}{(\xi-\xi_{0})^{2}+\eta^{2}}\\
    \notag&=\int d\xi\frac{f(\xi_{0})}{(\xi-\xi_{0})^{2}+\eta^{2}}
    +\int d\xi\frac{f(\xi)
    -f(\xi_{0})}{(\xi-\xi_{0})^{2}+\eta^{2}}\\
    &\longrightarrow\frac{\pi}{|\eta|}f(\xi_{0})+
    \lim_{\eta\rightarrow0^{+}}\int d\xi\frac{f(\xi)-f(\xi_{0})}{(\xi-\xi_{0})^{2}+\eta^{2}}
\end{align}
where the last line is in the limit of small $\eta$. The first term
corresponds to the transitions on resonance (or "on-shell"), where
energy is conserved in each tunneling event. These processes give
rise to a~current which is linear in $\Gamma$ after inserting into
Eq.~(\ref{eq:CurCotTot}) and using that $\eta\propto\Gamma$.
\textit{We can therefore omit the first term, since it has already
been calculated within the much simpler master equation scheme in
Sec.~\ref{sec:seq}}. The remaining term now corresponds to the
proper regularized second order contribution to the current. At this
point it is important to realize that this procedure does not
capture all second order terms, because the renormalization of the
dot state due to tunneling is not included.\cite{PhysRevB.54.16820}
The effect of renormalization, however, is easily incorporated by
adding these terms (linear in $\Gamma$) to the energies of the dot
states when doing the master equations. The second order correction
(in $\Gamma$) can then be extracted from the master equation result.
Here we do not incorporated them since they are unimportant and
merely give a shift of the gate voltage. On the other hand, for some
cases, \emph{e.g.} non-collinear magnetization, they cannot be
neglected because they give rise to off-diagonal elements in the
density matrix.\cite{PhysRevLett.90.166602}

Now returning to the expression \eqref{eq:PartSeqCo} we can evaluate
it analytically by employing the useful identity
\begin{equation}
    \lim_{\eta\rightarrow 0^{+}}\int d\xi\frac{f(\xi)-f(\xi_{0})}
    {(\xi-\xi_{0})^{2}+\eta^{2}}
    =\lim_{\eta\rightarrow 0^{+}}\frac{\partial}{\partial\xi_{0}}
    \mathrm{Re}\int d\xi\frac{f(\xi)}{\xi-\xi_{0}+i\eta}
\end{equation}
and summing over the residues. The final result becomes
\begin{align}
    \int_{-\infty}^{\infty}&\frac{d\xi}{(\xi-\xi_{0})^{2}}
    \big(n(\xi-\mu_{i})-n(\xi-\mu_{j})\big) \notag \\
    &\longrightarrow\textrm{Re}\bigg[\frac{\beta}{2\pi i}
    \Big(\Psi_{1}(\xi_{0},\mu_{i})
    -\Psi_{1}(\xi_{0},\mu_{j})\Big)\bigg].
\end{align}
The arrow indicates that the divergent integral has been regularized
by the procedure explained above. For convenience we introduced
a~shorthand notation for the polygamma function of n-$th$ order
\begin{equation}
    \Psi_{n}(\xi,\mu)\equiv
    \Psi_{n}\Big(\frac{1}{2}-\frac{\beta}{2\pi i}(\xi-\mu)\Big).
\end{equation}

Below we list the non-spin-flip currents, Eqs.~(\ref{eq:PairedCur})
after the regularization (for more detailed derivation of
regularized formulas see Ref.~\onlinecite{Zak07})
\begin{subequations}\label{eq:PairedCurReg}
\begin{align}
    \notag \tilde{J}_{ij}^{0}=&-\frac{e}{h}
    \Gamma^{0}_{i}\Gamma^{0}_{j}\textrm{Re}\big[
    \frac{\beta}{2\pi i}
    \big(\Psi_{1}(\epsilon_{\uparrow},\mu_{i})
    -\Psi_{1}(\epsilon_{\uparrow},\mu_{j})\big)\\
    &+\frac{\beta}{2\pi i}
    \big(\Psi_{1}(\epsilon_{\downarrow},\mu_{i})
    -\Psi_{1}(\epsilon_{\downarrow},\mu_{j})\big)\big],
\end{align}
\begin{align}
    \notag \tilde{J}_{ij}^{\sigma}=&-\frac{e}{h}
    \Gamma^{0}_{i}\Gamma^{0}_{j}\textrm{Re}\big[
    \frac{\beta}{2\pi i}
    \big(\Psi_{1}(\epsilon_{\sigma},\mu_{i})
    -\Psi_{1}(\epsilon_{\sigma},\mu_{j})\big)\\
    &+\frac{\beta}{2\pi i}
    \big(\Psi_{1}(\epsilon_{\bar{\sigma}}+U,\mu_{i})
    -\Psi_{1}(\epsilon_{\bar{\sigma}}+U,\mu_{j})\big)\big],
\end{align}
\begin{align}
    \notag \hspace{-2pt}\tilde{J}_{ij}^{2}=&-\frac{e}{h}
    \Gamma^{0}_{i}\Gamma^{0}_{j}\textrm{Re}\big[
    \frac{\beta}{2\pi i}
    \big(\Psi_{1}(\epsilon_{\uparrow}+U,\mu_{i})
    -\Psi_{1}(\epsilon_{\uparrow}+U,\mu_{j})\big)\\
    &+\frac{\beta}{2\pi i}
    \big(\Psi_{1}(\epsilon_{\downarrow}+U,\mu_{i})
    -\Psi_{1}(\epsilon_{\downarrow}+U,\mu_{j})\big)\big].
\end{align}
\end{subequations}

The current in the presence of the inelastic scattering,
Eq.~(\ref{eq:PairedCurSFDef}), causes more problems as the
interference term emerges. Using the partial fraction decomposition
it turns to be
\begin{align}
    \notag\bigg(&\frac{1}{\xi-\xi_{1}}-\frac{1}{\xi-\xi_{2}}\bigg)^{2}=\\
    &=\frac{1}{(\xi-\xi_{1})^{2}}+\frac{1}{(\xi-\xi_{2})^{2}}
    -\frac{2}{\xi_{1}-\xi_{2}}\bigg(
    \frac{1}{\xi-\xi_{1}}-\frac{1}{\xi-\xi_{2}}\bigg)
\end{align}
with the new type of divergence $(\xi-\xi_{0})^{-1}$. The identity
\begin{equation}
    \lim_{\eta\rightarrow 0^{+}}\int d\xi
    \frac{g(\xi)-g(\xi_{1})}
    {(\xi-\xi_{1})^{2}+\eta^{2}}
    =\lim_{\eta\rightarrow 0^{+}}
    \textrm{Re}\int d\xi\frac{f(\xi)}{\xi-\xi_{1}+i\eta},
\end{equation}
where
$g(\xi)\equiv(\xi-\xi_{1}+\frac{2\eta^{2}}{\xi_{1}-\xi_{2}})f(\xi)$,
allows one to find the regularized expression for this type of
divergence as well
\begin{align}
    \notag\int_{-\infty}^{\infty}&\frac{d\xi}{\xi-\xi_{0}}
    \big(n(\xi-\mu_{i})-n(\xi-\mu_{j})\big) \\
    &\longrightarrow\textrm{Re}\Big[\Psi_{0}\Big((\xi_{0},\mu_{j})
    -\Psi_{0}(\xi_{0},\mu_{i})\Big)\Big]
\end{align}
and after some algebra the inelastic tunneling rates become
\begin{align}
    \notag\Gamma_{j\bar{\sigma}i\sigma}=\frac{1}{h}
    &\Gamma^{0}_{i\sigma}\Gamma^{0}_{j\bar{\sigma}}
    n_{B}(\mu_{j}+\Delta_{\bar{\sigma}\sigma}-\mu_{i})\\
    \notag\times\textrm{Re}\Big[&\frac{\beta}{2\pi i}
    \big(\Psi_{1}(\epsilon_{\bar{\sigma}}+U,\mu_{j}+\Delta_{\bar{\sigma}\sigma})
    -\Psi_{1}(\epsilon_{\bar{\sigma}}+U,\mu_{i})\big)\\
    \notag+&\frac{\beta}{2\pi i}
    \big(\Psi_{1}(\epsilon_{\bar{\sigma}},\mu_{j}+\Delta_{\bar{\sigma}\sigma})
    -\Psi_{1}(\epsilon_{\bar{\sigma}},\mu_{i})\big)\\
    \notag+&\frac{2}{U}
    \big(\Psi_{0}(\epsilon_{\bar{\sigma}}+U,\mu_{j}+\Delta_{\bar{\sigma}\sigma})
    -\Psi_{0}(\epsilon_{\bar{\sigma}}+U,\mu_{i})\big)\\
    -&\frac{2}{U}
    \big(\Psi_{0}(\epsilon_{\bar{\sigma}},\mu_{j}+\Delta_{\bar{\sigma}\sigma})
    -\Psi_{0}(\epsilon_{\bar{\sigma}},\mu_{i})\big)\Big],
\end{align}
which substituted to Eq.~(\ref{eq:PairedCurSFDef}) gives a~proper
limit of non-interacting electrons,
$\lim_{U\rightarrow0}\tilde{J}^{\sigma,sf}_{ij}=0$. Note, that
current between each set of leads (involving spin's reversal or not)
fulfils the principle of detailed balance (vanishes for
$\mu_{i}=\mu_{j}$) and, hence, the total cotunneling current is
equal to zero, when the chemical potentials of the leads are at the
same level.

\section{\label{sec:sum}RESULTS}

\begin{figure}
\includegraphics[width=.45\textwidth]{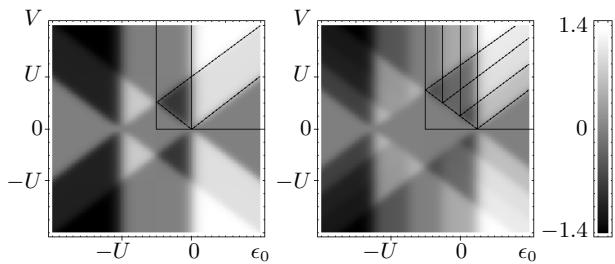}
\caption{\label{fig:Dens}
    The stem current in function of the bias voltage $V$ and the gate voltage
    $\epsilon_{0}$. The leads are unpolarized and equally coupled to the dot,
    with coupling's strength $\Gamma/3$. The picture without the magnetic field
    (on the left) differs from this with field $B=5\Gamma$ (on the right).
    The interaction energy $U=20\Gamma$ and temperature $k_{B}T=\Gamma$.
    The grey scale in the far right describes the stem current value in units
    $e\Gamma/\hbar$.}
\end{figure}

In Fig.~\ref{fig:Dens} the stem current dependence upon the bias
voltage $V$ and the gate voltage $\epsilon_{0}$ for no magnetic
field (the left plot) and for magnetic field $B=5\Gamma$ (right) is
shown. Bright regions correspond to positive values of the current
while dark areas to negative ones. It is apparent that the stem
current is an~even function of bias $V$ and an~odd function of the
orbital energy $\epsilon_{0}$ with respect to the particle-hole
symmetry line $\epsilon_{0}=-U/2$. Therefore, we restrict our
discussion to the parts of the plots where $V>0$ and
$\epsilon_{0}>-U/2$. By performing the particle-hole transformation,
the behavior in region where $\epsilon_{0}>-U/2$ holds, can be
mapped onto the remaining area of the $V-\epsilon_{0}$ space.

Consider the case without the magnetic field first. If
$\epsilon_{0}<0$, the dot is in the single occupied state, and until
$V<\epsilon_{0}$ the transport is blocked because electrons from the
left branch do not have enough energy to overcome the Coulomb
blockade. Increasing voltage makes the transport out of the stem
possible, since electrons may escape to the empty states in the
right lead. However, further increase of the voltage stops the stem
current again, because for $V>\epsilon_{0}+U$ (when two electrons
excitations become possible) the left reservoir supplies the stem
with electrons while at the same time the electrons from the stem
move into the right reservoir. These currents cancel one another and
there is no net stem current. The situation differs for
$\epsilon_{0}>0$. The current increases once $V>\epsilon_{0}$, when
dot is excited to the single occupied state, and again when
$V>\epsilon_{0}+U$, where the electrons transverse through the
double occupied state. In both cases the direction of the flux is
directed into the stem and carried by the left-lead electrons.
Clearly, for the negative voltages the role of the right and left
reservoir interchanges, but nevertheless the direction of the stem
current is unaffected.

In the presence of the magnetic field more complex structure emerges
due to two new excitations coming into play (as the single occupied
state is no longer degenerate). It is convenient to divide the right
part of the diagram into four vertical strips
$\epsilon_{0}\in[-U/2,-U/2+B]$, $[-U/2+B,0]$, $[0,B]$ and
$\epsilon_{0}>B$.

In the first region, the negative stem current flows in the narrow
range of voltages. Below the lower threshold $V<-\epsilon_{0}+B$,
the spin-up electron occupying the dot stops the current until the
required state in the right reservoir becomes available. Then, both
spin-up and spin-down electrons can participate in transport. For
$V>\epsilon_{0}+U+B$ the electrons from the left branch enter and
compensate the current carried from the stem to the right lead and,
hence, the net stem current vanishes.

\begin{figure}
\includegraphics[width=.45\textwidth]{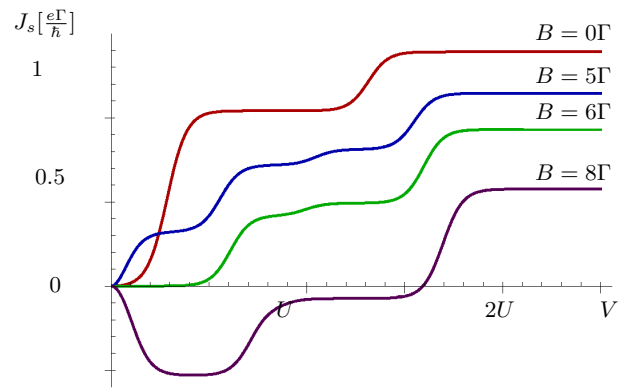}
\caption{\label{fig:JsvsB}(Color online) The stem current for
    different values of magnetic field $B$ is plotted. Interestingly,
    there is a~sign change in the vicinity of $V=0$ for $B=\epsilon_{0}$.
    See text for explanation. The gate voltage is $\epsilon_{0}=6\Gamma$.
    Other parameters are as in the previous figure.}
\end{figure}

In the second region, from $V=0$ to $V=-\epsilon_{0}+B$ the dot is
occupied by the spin-up electron which blocks the current. Beyond
$V=-\epsilon_{0}+B$ the spin-up states in the right lead become
available and this electron moves towards the right reservoir.
Further increase of the bias voltage results in a~small increase of
the current at $V=\epsilon_{0}+U-B$, when spin-up electrons from the
left electrode can pass through the Coulomb blockade. Finally, there
is a~sign change of the stem current at $V>\epsilon_{0}+U+B$,
because electrons of both spin directions go into the stem from the
left lead while still only spin-up electrons can move into the right
reservoir.

The negative stem current in the third strip starts to flow when
$V>-\epsilon_{0}+B$ due to the spin-up electrons moving out of the
stem into the right reservoir. This current is risen by the
spin-down electrons from the left lead at $V=\epsilon_{0}+B$ and
also by electrons moving through the double occupied state (for both
$V=\epsilon_{0}+U-B$ and $V=\epsilon_{0}+U+B$).

Eventually, in the last strip the stem current has four steps. These
are for $V=\epsilon_{0}-B$, $V=\epsilon_{0}+B$ when the excitations
of the empty dot to spin-up and spin-down state respectively become
energetically allowed, and for $V=\epsilon_{0}+U-B$,
$V=\epsilon_{0}+U+B$ when electrons from the right lead have enough
energy to overcome the Coulomb blockade (due to the spin-down and
then spin-up electron on the dot).

\begin{figure}
\includegraphics[width=.45\textwidth]{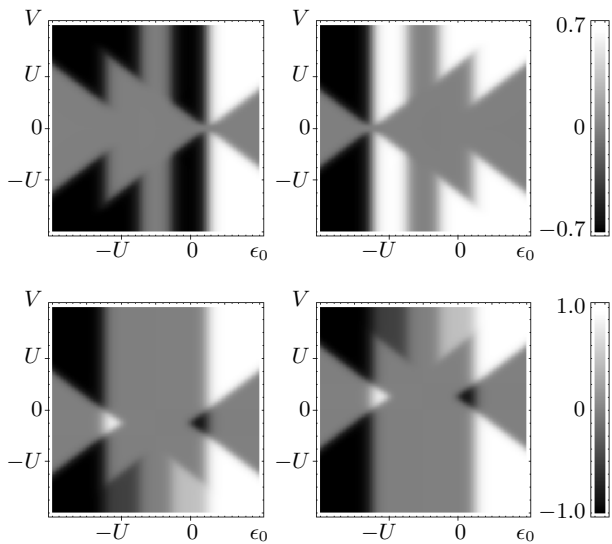}
\caption{\label{fig:DensB}
    The stem current for different polarizations of leads and magnetic field
    $B=5\Gamma$. The central branch is always unpolarized.
    In the upper row, both left and right leads are spin-up polarized,
    $P_{l}=P_{r}=1$ (the left plot), or spin-down polarized, $P_{l}=P_{r}=-1$
    (the right plot). Due to nonzero magnetic field, the spin symmetry is broken
    and the stem current is no longer an~odd function of the gate voltage with respect to
    $\epsilon_{0}=-U/2$. In the lower row, polarization of the left and right
    leads is antiparallel ($P_{l}=1$, $P_{r}=-1$ on the left, and $P_{l}=-1$,
    $P_{r}=1$ on the right). In this case, the current symmetry in $V$ is violated,
    because the dot couples to the left and right leads asymmetrically.
    The other parameters are as in the previous figure.}
\end{figure}

Fig.~\ref{fig:JsvsB} shows how the current is affected by the
magnetic field applied to the dot. The stem current remains positive
as long as $B\leq\epsilon_{0}$ for any bias. The character of the
function changes when $B$ crosses $\epsilon_{0}$. This corresponds
to the situation when the single occupancy state has the same energy
as the empty state (the electron-hole symmetry point). Tuning the
magnetic field to $B=\epsilon_{0}$ enables one to measure directly
electron-hole transport fluctuations.

Providing $B>\epsilon_{0}$, the spin-up state of the dot is below
the Fermi level of the stem and for small voltages the negative
current is most likely to occur. However, the increasing voltage
forces more and more electrons to move out of the left lead into the
central branch. These electrons overbalance those heading in the
opposite direction and the current sign changes -- it is apparent
for $B=8\Gamma$.

In two-lead systems the tunneling magnetoresistance (TMR) have
widely been
studied.\cite{PhysRevB.62.1186,PhysRevB.64.085318,weymann:115334}
Defining a~similar quantity in the multiprobe setup is rather vague.
Nonetheless, it turns out that the current through the central,
unpolarized lead has interesting features while varying
magnetization of the other electrodes. The results are shown in
Fig.~\ref{fig:DensB}.

The parallel magnetization (the upper part) destroys the
antisymmetry in the gate voltage whereas the antiparallel alignment
(the lower part) breaks the bias voltage symmetry. It is worth to
note that the third lead gives not only information about the
relative magnetization of two remaining electrodes, but also allows
one to determine polarization of each lead. This means that in that
kind of a~device we can switch among four different states and
directly read out the information decoded in the two-lead system
with the third probe. Experimentally it might be more convenient to
swap the magnetic field on the dot rather than to change
polarization of the side leads.

\section{\label{sec:conclu}CONCLUSIONS}

In conclusion, we have studied Coulomb blockade in a~three terminal
device. We have focused on the bias situation where two leads
have opposite voltage, while the third lead (the stem) is grounded.
This setup allows for a~direct measure of the electron-hole
asymmetry of the quantum dot system and we have made detailed
calculations of the stem current in the gate voltage-bias voltage plane.
We have pointed to a~number of predictions, that can be
experimentally tested.

Furthermore, we have considered the spin polarized case, where
a~number of detailed experimental proposals have been presented.
In particular, a~large difference in predicted pattern between
parallel and anti-parallel configurations is seen.

Finally, we have checked, using a~regularized version of the usual
cotunneling formalism, that our predictions are not significantly
altered by cotunneling corrections.

\appendix*
\section{\label{app:exc}NON-INTERACTING ELECTRONS LIMIT}

The limiting case of non-interacting electrons gives the opportunity
to compare the results we get within the framework of the second
order perturbation theory with the exact result containing higher
order terms in couplings. The quality of the cut-off on the second
order terms can be estimated. We derive the generalization of the
Meir--Wingreen formula that will include the indefinite number of
leads and calculate the current $J_{i}$ through one of the
junctions.

The main result of Meir and Wingreen\cite{Meir92} is the current
through lead $i$ in the presence of interactions
\begin{equation}\label{eq:CurrentMeirWingr}
    J_{i}=\frac{ie}{h}\int d\xi\bigg\{\textrm{Tr}
    \big[\Gamma_{i}(G^{R}-G^{A})\big]n(\xi-\mu_{i})
    +\textrm{Tr}\big[\Gamma_{i}G^{<}\big]\bigg\}.
\end{equation}
This can be generalized to the many lead systems employing the
obvious identity$J_{i}=\gamma J_{i}-(1-\gamma)\sum_{j\neq i}J_{j}$
following from Kirchhoff's law. Providing all the couplings are
proportionate, that is $\Gamma_{i}=\lambda_{ij}\Gamma_{j}$ where
$\lambda_{ij}$ are constants, one may eliminate the lesser Green
function $G^{<}$ by proper selection of $\gamma$ showing that
\begin{equation}\label{eq:CurrentMeirWingrGen}
    J_{i}=-\frac{e}{h}\int d\xi\sum_{j=1}^{N}
    \textrm{Tr}\bigg[A\frac{\Gamma_{i}\Gamma_{j}}{\Gamma}\bigg]
    \big(n(\xi-\mu_{i})-n(\xi-\mu_{j})\big),
\end{equation}
where we introduced a~shorthand notation
$\Gamma\equiv\sum_{i}\Gamma_{i}$ and wrote the formula for the
current $J_{i}$ in terms of a~spectral function $A\equiv
i(G^{R}-G^{A})$.

For non-interacting electrons
\begin{equation}
    A(\xi,\sigma)=
    \frac{\Gamma}{\big(\xi-\epsilon_{\sigma}\big)^{2}
    +\big(\frac{\Gamma}{2}\big)^{2}}
\end{equation}
\begin{figure}
\includegraphics[width=.45\textwidth]{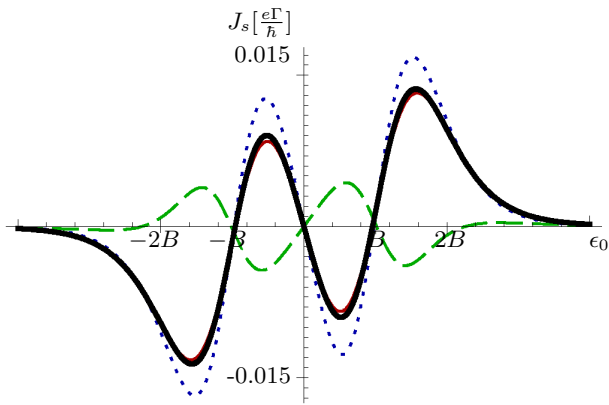}
\caption{\label{fig:NonIntB}(Color online)
    The strict result for the stem current (thick, black) compared to the
    perturbative result: sequential (dotted, blue), cotunneling (dashed,
    green), and their sum (thin, red). Parameters are as in Fig.~(\ref{fig:Jtot}) (except for $U=0$).}
\end{figure}describes a~spin dependent spectral function, which substituted to
the generalized Meir-Wingreen formula gives the exact current in
lead $i$ for the multiterminal device, valid to any order in
couplings
\begin{equation}\label{eq:CurrentLanButtS}
    J_{i}=-\frac{e}{h}\Gamma_{i}
    \sum_{j}\sum_{\sigma}\Gamma_{j}\int d\xi
    \frac{n(\xi-\mu_{i})-n(\xi-\mu_{j})}
    {\big(\xi-\epsilon_{\sigma}\big)^{2}
    +\big(\frac{\Gamma}{2}\big)^{2}}.
\end{equation}
We note in passing that the same result comes from the
non-interacting Landauer--B\"utticker formalism for any number of
electrodes.

Using the similar methods as described in Sec.~\ref{sec:cotreg}, we
derive the current through junction $i$ that depends on coupling
$\Gamma$ up to any order
\begin{equation}
    J_{i}=-\frac{4e}{h}\Gamma_{i}
    \sum_{j}\sum_{\sigma}\Gamma_{j}
    \textrm{Im}\bigg[\Psi_{0}(\epsilon_{\sigma},\mu_{i},\Gamma)
    -\Psi_{0}(\epsilon_{\sigma},\mu_{j},\Gamma)\bigg]
\end{equation}
and
\begin{equation}
    \Psi_{0}(\epsilon_{0},\mu_,\Gamma)
    \equiv\Psi_{0}\Big(\frac{1}{2}-\frac{\beta}{2\pi i}(\epsilon_{0}-\mu)
    +\frac{\beta}{4\pi}\Gamma\Big).
\end{equation}

In Fig.~\ref{fig:NonIntB} the exact result is compared with the
perturbative one. The good agreement between these two justifies the
choice of the method, and reassures that the second order
perturbation theory gives not only excellent quantitative
description, reachable also within the sequential tunneling
framework, but also good qualitative estimation, at least for
non-interacting electrons. Note, however, that this conclusion is
only valid in the regime where $\Gamma\ll k_{B}T$, whereas in the
opposite limit the perturbation expansion clearly fails, and one does
not expect the good agreement.

\begin{acknowledgments}
We acknowledge inspiring discussions with A. M. Lunde, J. A.
Majewski, J. Paaske, M. Wegewijs. R.~A.~\.Zak appreciates the
hospitality of the Niels Bohr Institute, University of Copenhagen,
and the financial support from the European Physical Society.
\end{acknowledgments}


\end{document}